\documentclass[12pt,preprint]{aastex}

\slugcomment{The Astrophysical Journal, Submitted 2004 January}

\shortauthors{Itoh et al.}
\shorttitle{ION-ION CORRELATION EFFECT ON THE NEUTRINO-NUCLEUS SCATTERING}

\received{2004 January 20}
\begin{document}

\title{ION-ION CORRELATION EFFECT ON THE NEUTRINO-NUCLEUS SCATTERING IN SUPERNOVA CORES}

\author{\sc Naoki Itoh, Ryohei Asahara, Nami Tomizawa, and Shinya Wanajo}
\affil{Department of Physics, Sophia University,
       7-1 Kioi-cho, Chiyoda-ku, Tokyo, 102-8554, Japan;\\
       n\_itoh@sophia.ac.jp, r-asahar@sophia.ac.jp, tomiza-n@sophia.ac.jp, 
       wanajo@sophia.ac.jp}

\and

\author{\sc Satoshi Nozawa}
\affil{Josai Junior College for Women, 1-1 Keyakidai, Sakado-shi,
       Saitama, 350-0295, Japan; snozawa@josai.ac.jp}

\bigskip
\affil{The Astrophysical Journal, Submitted 2004 January}

\begin{abstract}

We calculate the ion-ion correlation effect on the neutrino-nucleus scattering in supernova cores, which is an important opacity source for the neutrinos and plays a vital role in the supernova explosion.  In order to calculate the ion-ion correlation effect we use the results of the improved hypernetted-chain method calculations of the classical one-component plasma.  As in the preceding studies on this effect, we find a dramatic decrease of the effective neutrino-nucleus scattering cross section for relatively low energy neutrinos with $E_{\nu} \leq 20$ MeV.  As a matter of fact, our calculation shows a much more dramatic reduction of the effective neutrino-nucleus scattering cross section for the low energy neutrinos with $E_{\nu} \leq 10$ MeV than the results of Horowitz.  Therefore, the ion-ion correlation effect will be more important than has hitherto been recognized.  We present an accurate analytic fitting formula that summarizes our numerical results.  This fitting formula will facilitate the application of the present results to the supernova explosion simulations.

\end{abstract}


\section{INTRODUCTION}

In a recent important paper, Liebend\"orfer et al. (2003) have made thorough comparisons of various neutrino radiation-hydrodynamical codes of the supernova explosion caused by the delayed neutrino-driven mechanism.  They have carefully assessed the magnitude of the ion-ion correlation effect on the neutrino-nucleus scattering in the supernova cores and have found that this effect causes an appreciable difference in the supernova core collapse results.

Historically, Itoh (1975) pointed out the importance of the ion-ion correlation effect on the neutrino-nucleus scattering in the supernova cores for the first time shortly after Freedman's (1974) theoretical discovery of the coherent aspect of the neutrino-nucleus scattering and its importance to the supernova problem.  Itoh (1975) used Hansen's (1973) rather limited Monte Carlo results on the classical one-component plasma then available and showed the appearance of the dramatic reduction of the effective neutrino-nucleus scattering cross section caused by the ion-ion correlation effect for relatively low energy neutrinos.  Bowers \& Wilson (1982) subsequently published an analytic fitting formula for the ion-ion correlation effect that is based on the work of Bond (1980).  Bruenn (1993) discussed the consequence of the ion-ion correlation effect on the neutrino-nucleus scattering in supernova cores in great detail.

More recently Horowitz (1997) presented a detailed fitting formula for the ion-ion correlation effect based on his Monte Carlo calculations.  Bruenn \& Mezzacappa (1997) and Liebend\"orfer et al. (2003) have made a detailed assessment of the importance of the ion-ion correlation effect on the neutrino-nucleus scattering in the supernova explosion using the results of Horowitz (1997).  Since this effect plays an important role in the supernova explosion, we wish to carry out an independent calculation of the ion-ion correlation effect on the neutrino-nucleus scattering and make a detailed comparison with Horowitz's results in this paper.

The present paper is organized as follows.  Physical conditions relevant to the present calculation are made explicit in \S~2.  Calculation of the ion-ion correlation effect on the neutrino-nucleus scattering in supernova cores is presented in \S~3.  An analytic fitting formula that summarizes the numerical results is presented in \S~4.  Concluding remarks are given in \S~5.

\section{PHYSICAL CONDITIONS}

We assume that a supernova core consists of atomic nuclei $(Z,A)$ and degenerate electrons.  At the densities and temperatures relevant to the supernova core, these nuclei interact with each other strongly by the Coulomb interaction.  The conventional parameter that characterizes such strongly-interacting classical ions is (Itoh et al. 2003)
\begin{eqnarray}
\Gamma & = & \frac{ \left(Ze \right)^{2} }{a_{I} k_{B} T} \, = \,
0.2275 \, \frac{Z^{2}}{T_{10}} \left( \frac{ \rho_{12}}{A} \right)^{1/3} \, , 
\end{eqnarray}
where $a_{I}$ is the ion-sphere radius
\begin{eqnarray}
a_{I} & = & 0.7346 \times 10^{-12} \left( \frac{ \rho_{12}}{A} \right)^{-1/3} \,  {\rm cm} \, ,
\end{eqnarray}
$Z$ being the atomic number of the nucleus, $A$ being the mass number of the nucleus, $T_{10}$ being the temperature in units of 10$^{10}$K, and $\rho_{12}$ being the mass density in units of 10$^{12}$ g cm$^{-3}$.

In this paper we restrict ourselves to the case in which electrons are strongly degenerate.  This condition is expressed as
\begin{eqnarray}
T \ll T_{F} & = & 5.930 \times 10^{9} \, \left[ \left\{ 1 + 1.018 
 (Z/A)^{2/3} \rho_{6}^{2/3} \right\}^{1/2} - 1 \right] \,  {\rm K} \, ,
\end{eqnarray}
where $T_{F}$ is the electron Fermi temperature and $\rho_{6}$ is the mass density in units of 10$^{6}$ g cm$^{-3}$.  We impose a further condition that the ions are in the liquid state (Nagara, Nagata, \& Nakamura 1987),
\begin{equation}
\Gamma < 172 \, .
\end{equation}

The conventional $r_{s}$-parameter for the electron liquid is given by (Itoh et al. 2002)
\begin{eqnarray}
 r_{s} & = & \frac{ a_{e} }{ \hbar^{2}/m_{e}e^{2}} = 1.388 \times 10^{-4} 
             \left( \frac{A}{Z \rho_{12}} \right)^{1/3} \, ,   \\
  &  &  \frac{4}{3} \pi a_{e}^{3} \, n_{e} \, = \, 1  \, ,
\end{eqnarray}
where $n_{e}$ is the electron number density.  The electron Fermi wave number
is expressed as
\begin{eqnarray}
k_{F} & = & 2.613 \times 10^{12} \, \left( \frac{Z}{A} \, \rho_{12} 
            \right)^{1/3} \, \, {\rm cm}^{-1} \, .
\end{eqnarray}
We have the relationship
\begin{eqnarray}
k_{F} a_{I} & = & \left( \frac{ 9 \pi}{4} \right)^{1/3} \, Z^{1/3} \, .
\end{eqnarray}

When electrons are extremely relativistic, the Fermi-Thomas screening wave number is expressed as (Flowers \& Itoh 1976)
\begin{eqnarray}
\frac{k_{FT}}{k_{F}} & \simeq & \left( \frac{4}{ \pi} \frac{1}{137.036} \right)^{1/2} \, = \, 0.0964 \, .
\end{eqnarray}
Therefore, for the extremely relativistic electrons, one obtains
\begin{eqnarray}
k_{FT} a_{I} & = & 0.1850 \, Z^{1/3} \, .
\end{eqnarray}
In Figure~1 we show the density-temperature diagram of the pure $^{56}$Fe plasma that illustrates the physical conditions described in this section.  Here one should note that we assume the ion-ion correlation effect is caused only by the Coulomb interaction between the nuclei.  At extremely high densities $\rho \gtrsim 10^{14}$ g cm$^{-3}$ one should also take into account the nuclear (strong) interaction to calculate the ion-ion correlation effect accurately.

\section{ION-ION CORRELATION EFFECT}

Bruenn \& Mezzacappa (1997) have given detailed expressions for the ion-ion correlation effect on the neutrino-nucleus scattering in the supernova core.  We refer the reader to this very comprehensive paper for the detailed discussion of this effect.

It is well-known in liquid metal physics that one has to include the liquid structure factor to obtain the effective cross section for the X-ray scattering in the liquid metals (March 1968).  This condition equally applies to the neutrino-nucleus scattering in supernova cores, as the nuclei are strongly correlated with each other by the Coulomb interaction.  The strength of the ion-ion correlation is measured by the Coulomb coupling parameter $\Gamma$ in equation (1).

The effective transport cross section for the neutrino-nucleus scattering is given by (Freedman 1974; Itoh 1975)
\begin{eqnarray}
\left< \frac{d \, \sigma}{d \, {\rm cos} \theta} \right> & = & a_{0}^{2} \frac{G^{2}}{2 \pi} A^{2} E_{\nu}^{2} \left(1 + {\rm cos} \theta \right) \left(1 - {\rm cos} \theta \right) S(k)  \, ,
\end{eqnarray}
where $\theta$ is the neutrino laboratory scattering angle, $a_{0}$ is a numerical parameter of the theory of the weak interaction, $G$ is the Fermi coupling constant, $A$ is the mass number of the nucleus, and $S(k)$ is the liquid structure factor.  The momentum transfer $k$ is related to the neutrino energy $E_{\nu}$ by 
\begin{eqnarray}
k & = & \frac{2 E_{\nu}}{\hbar c} \, {\rm sin} \frac{\theta}{2} \, .
\end{eqnarray}
We refer the reader to Bruenn \& Mezzacappa (1997) for more detailed expressions for the cross section.

For the uncorrelated ions with $S(k)=1$, we have the angle-integrated cross section
\begin{eqnarray}
< \sigma >_{uncorr} & = & \frac{4}{3} a_{0}^{2} \frac{G^{2}}{2 \pi} A^{2} E_{\nu}^{2}  \, .
\end{eqnarray}
Therefore, we can express the ion-ion correlation effect $S(k) \neq 1$ by writing
\begin{eqnarray}
< \sigma >_{corr} & = & \frac{4}{3} a_{0}^{2} \frac{G^{2}}{2 \pi} A^{2} E_{\nu}^{2} < S > \, , \\
< S > & = & \frac{3}{4} \int d {\rm cos} \theta \left(1 + {\rm cos} \theta \right) \left(1 - {\rm cos} \theta \right) S(k)   \nonumber  \\
& = & 12 \int_{0}^{1} x^{3} (1 - x^{2}) S(x) d x \, , \\
x & \equiv & \frac{a_{I} k}{2 E_{\nu} a_{I}/\hbar c} \, = \, \frac{a_{I}k}{2 \epsilon} \, ,  \\
\epsilon & \equiv & \frac{E_{\nu} a_{I}}{\hbar c} \, = \, \frac{E_{\nu} a_{I}}{197.3 \, {\rm MeV \, fm}} \, .
\end{eqnarray}

So far we have considered only one nuclear species $(Z,A)$.  However, in the real supernova explosion more than one nuclear species exist in supernova cores (Liebend\"orfer et al. 2003).  Here we give a prescription to calculate the neutrino mean free path $\ell$ due to the neutrino-nucleus scattering including protons:
\begin{eqnarray}
\ell^{-1} & = & \sum_{j} n_{j} < \sigma_{j} > \, ,
\end{eqnarray}
where $n_{j}$ is the number density of ions of $j$-th species and $< \sigma_{j} >$ is the neutrino-nucleus scattering cross section by $j$-th species ions (nuclei).  In calculating $< \sigma_{j} >$ for the neutrino-nucleus scattering by $j$-th nuclear species $(Z_{j},A_{j})$, one uses the mass number $A_{j}$ and the angle-averaged ion-ion correlation factor $<S>$ in equation (14) corresponding to the $\Gamma$-value (Itoh et al. 1979)
\begin{eqnarray}
\Gamma_{j} & = & \frac{Z_{j}^{5/3} e^{2}}{a_{e} k_{B} T} \, = \, 0.2275 \, \frac{Z_{j}^{5/3}}{T_{10}} \left( \rho_{12} \sum_{i} \frac{X_{i}Z_{i}}{A_{i}} \right)^{1/3} \, ,
\end{eqnarray}
where $X_{i}$ is the mass fraction of the $i$-th nuclear species $(Z_{i},A_{i})$ and $a_{e}$ is the electron-sphere radius
\begin{eqnarray}
a_{e} & = & \left( \frac{3}{4 \pi n_{e}} \right)^{1/3} = \left( \frac{3}{4 \pi \sum_{i} n_{i} Z_{i}} \right)^{1/3} \, ,
\end{eqnarray}
$n_{e}$ being the electron number density.

For the $^{56}$Fe plasma with $Z$=26, we have $k_{FT} a_{I}$=0.548 from equation (10).  Hubbard \& Slattery (1971) have carried out a Monte Carlo simulation of the ions screened with the dielectric function of the electron liquid.  They have shown that for $k_{FT} a_{I} \lesssim 0.5$ the results that take into account the electron dielectric function are close to the results corresponding to the ions imbedded in the negative background.  Therefore, we are justified in using the liquid structure factor corresponding to the ions imbedded in the negative background (classical one-component plasma).

In this paper we use the liquid structure factor of the classical one-component plasma calculated by Ichimaru, Iyetomi, \& Tanaka (1987) using the improved hypernetted-chain scheme.  They have presented a numerical table for $\Gamma$=2, 5, 10, 20, 40, 80, 125, and 160.  We supplement this table by the table  presented in Itoh et al. (1983) by using the same method for the case of $\Gamma$=1.  In passing, it should be noted that those who wish to calculate the angle-dependent cross section should use the tables of the liquid structure factor $S(k)$ presented in Ichimaru, Iyetomi, \& Tanaka (1987) and also in Itoh et al. (1983).  Constructing analytic fitting formulae for $S(k)$ would be extremely difficult, since $S(k)$ has oscillatory behavior.  It would be best to use numerical tables rather than analytic fitting formulae for $S(k)$.

In Figure~2 we show the results of the calculation for $<S>$ in equation (15).  A similar graph has been presented in Bruenn \& Mezzacappa (1997).  They have used Horowitz's (1997) fitting formula.

In Figure~3 we compare our results with Horowitz's fitting formula.  We find that Horowitz's results for $\epsilon \leq 1.0$ greatly differ from our results. It appears that this is due to the inaccurate treatment done by Horowitz to calcuate the small-$k$ behavior of the liquid structure factor $S(k)$ that cannot be directly obtained from the Monte Carlo results because of the finite size of the simulation.  We have used the correct small-$k$ ($a_{I}k < 1$) behavior of the liquid structure factor (Ichimaru 1982; Itoh et al. 1983):
\begin{eqnarray}
S(k) & = & \left[ \frac{3 \Gamma}{(a_{I}k)^{2}} + \frac{1}{k_{B}T} \left(\frac{ \partial P}{\partial n} \right)_{T} \right]^{-1} \, , 
\end{eqnarray}
\begin{eqnarray}
\frac{1}{k_{B}T} \left(\frac{ \partial P}{\partial n} \right)_{T} & = & 0.73317 - 0.39890 \, \Gamma + 0.34141 \, \Gamma^{1/4} + 0.05484 \, \Gamma^{-1/4} \, .
\end{eqnarray}

We find that the present more accurate calculation leads to a more dramatic reduction of the neutrino-nucleus scattering cross section for $\epsilon \leq 1.0$ than has been calculated by Horowitz (1997).  Therefore, the ion-ion correlation effect on the neutrino-nucleus scattering in supernova cores will be more important than has hitherto been recognized.  However, the consequence of the modification in the ion-ion correlation effect may not be so great as to drastically change the whole scenario of the supernova explosion, since the phase space for these low energy neutrinos is relatively small.

\section{ANALYTIC FITTING FORMULA}

In this section we present an accurate analytic fitting formula for $<S>(\Gamma,\epsilon)$ in order to facilitate the application of the numerical results obtained in the present paper.  We have carried out the numerical calculations of $<S>(\Gamma,\epsilon)$ for $1 \leq \Gamma \leq 160$, $0.01 \leq \epsilon \leq 10.0$.  We express the analytic fitting formula for this range by
\begin{eqnarray}
<S>_{fit} &=& \frac{X(\gamma, \xi)}{\displaystyle{1 + \frac{3}{2} \frac{\Gamma}{\epsilon^{2}}                \exp \left(-\frac{\epsilon^2}{\Gamma^{1/16}} \right)}}   \, , \\
X(\gamma, \xi)         &=& 1 \, - \, \frac{1}{7.8}(\xi-1)^2(\gamma+2)^2 \,
                \exp\left\{-10\left(\xi-\frac{\gamma}{10}-0.25\right)^2
                 - 0.8(\gamma-0.2) \right\} \,  \nonumber \\ 
          & & \, \, \, \, \, - \, \frac{1}{10}(\gamma+2)^{-0.3} \,
                \exp\left\{-6\left(\xi-\frac{\gamma}{10}+0.2\right)^2\right\}
                 \nonumber \\ 
          & & \, \, \, \, \, + \, \frac{1}{74}(\gamma+1.1)^5
                \exp\left\{-130\left(\xi+\frac{\gamma}{10}-0.5\right)^2\right\}
                 \nonumber \\ 
          & & \, \, \, \, \, - \, \frac{1}{42.2}(\gamma+2)^2
                \exp\left\{-20(\gamma+2)(\xi-0.25)^2 - 0.8(\gamma-0.2)\right\} 
                 \, , \\
 \gamma & \equiv & \frac{1}{1.10206} \left( {\rm log}_{10} \Gamma \, - \, 1.10206 \right)  \, ,  \, \, \, \, {\rm for \, \, } 1 \leq \Gamma \leq 160  \\
 \xi & \equiv & {\rm log}_{10} \epsilon \, , \, \, \, \,  {\rm for \, \, } 0.01 \leq \epsilon \leq 10.0.
\end{eqnarray}
The ratio of the present fitting formula value to our numerical result $<S>_{fit}/<S>$ is shown in Figure~4.  It is readily seen that the accuracy of the fitting is better than 6\% for all the parameter range considered.

\section{CONCLUDING REMARKS}

We have calculated the ion-ion correlation effect on the neutrino-nucleus scattering in supernova cores by using the liquid structure factor of the classical one-component plasma obtained by the improved hypernetted-chain scheme.  We have found a dramatic reduction of the neutrino-nucleus scattering cross section due to the ion-ion correlation effect.  We have compared the results of our calculation with those of Horowitz.  We have found Horowitz's calculation greatly differs from ours for $\epsilon \equiv E_{\nu}a_{I}/\hbar c \leq 1$ because of his inaccurate treatment for the calculation of the liquid structure factor $S(k)$ for small-$k$ values.  Thererfore, we have found that the ion-ion correlation effect is more important than has hitherto been recognized.  We have presented an accurate analytic fitting formula that summarizes our numerical results to facilitate the application.  Supernova collapse simulations that take into account the present results are strongly awaited.

\acknowledgments

We wish to thank N. W. Ashcroft, G. Chabrier, W. B. Hubbard, T. Janka and M. Liebend\"orfer for most helpful communications.  We also wish to thank our referee for very useful comments that helped us improve the paper.  This work is financially supported in part by Grants-in-Aid of the Japanese Ministry of Education, Culture, Sports, Science, and Technology under contracts \#13640245 and \#15540293.

\clearpage

\begin{figure}
\epsscale{0.7}
\plotone{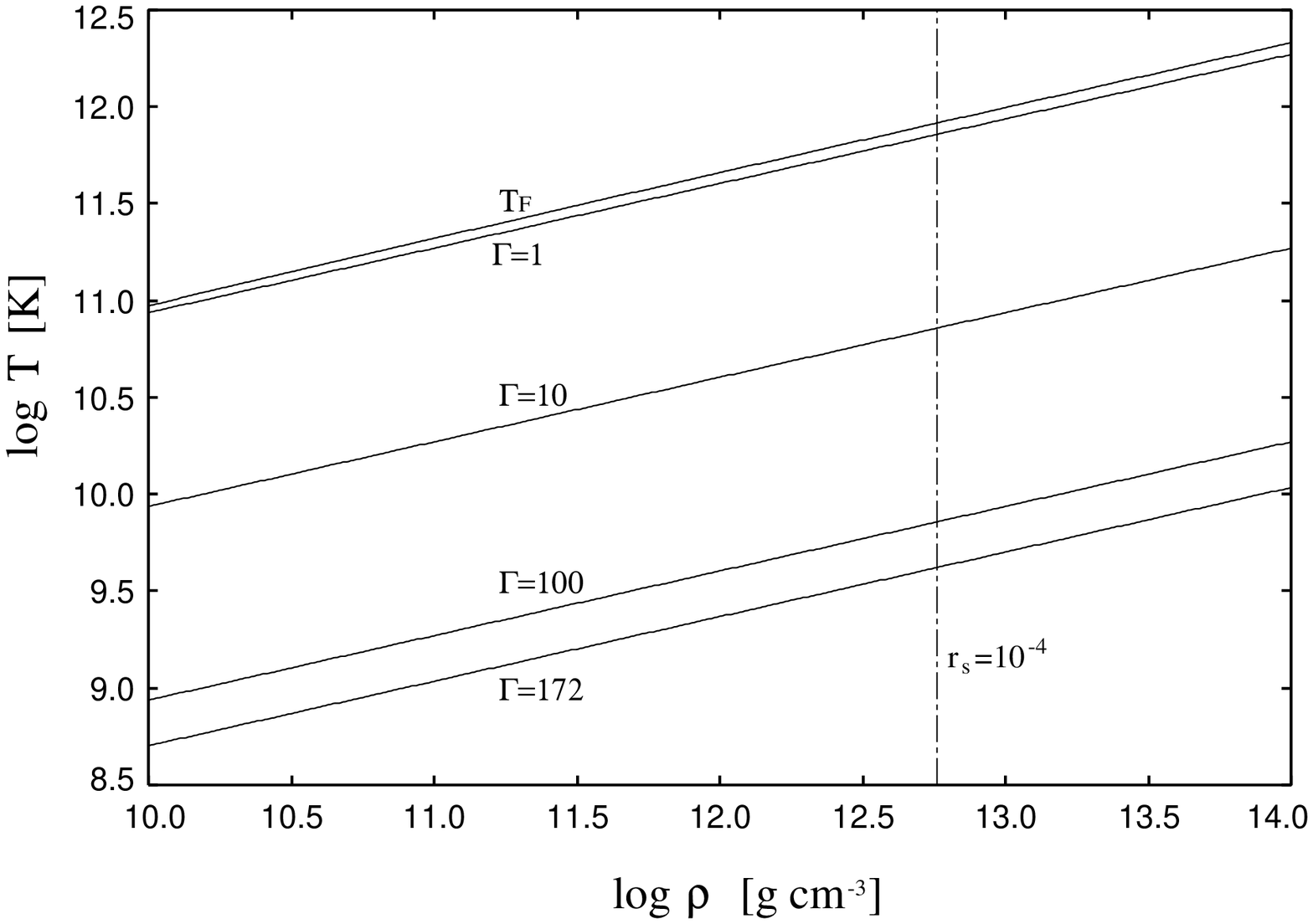}
\caption{Physical conditions for the pure $^{56}$Fe plasma.  The present calculation is valid below the electron Fermi temperature $T_{F}$.  Furthermore, the ions should be in the liquid state, $\Gamma < 172$.  The lines of $\Gamma$=constant and $r_{s}$=constant are also shown.}
\end{figure}

\begin{figure}
\epsscale{0.7}
\plotone{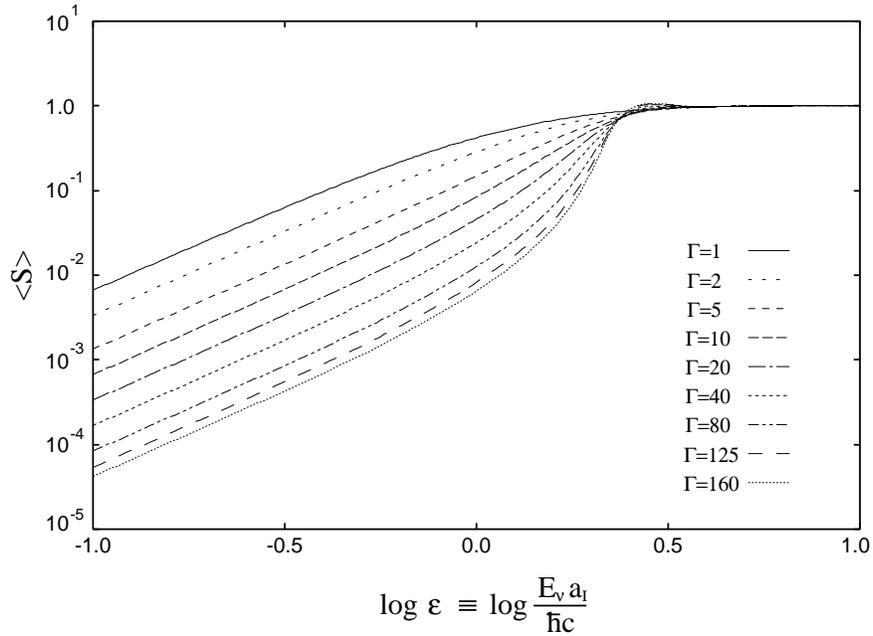}
\caption{The angle-averaged ion-ion correlation factor $<S>$ as a function of $\epsilon \equiv E_{\nu} a_{I}/ \hbar c$ for various values of $\Gamma$.}
\end{figure}

\begin{figure}
\epsscale{0.7}
\plotone{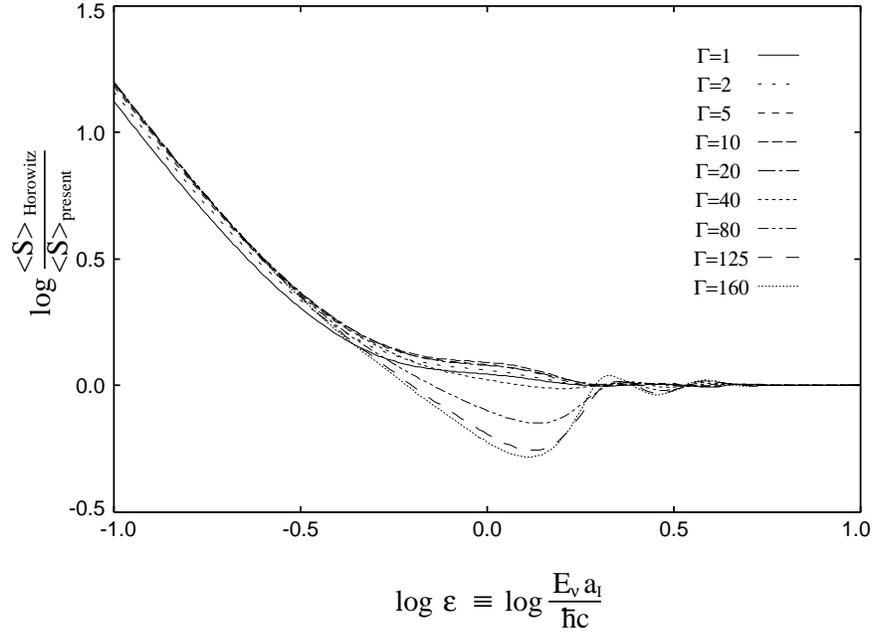}
\caption{Comparison of the present calculation with Horowitz's fitting formula.}
\end{figure}

\begin{figure}
\epsscale{0.7}
\plotone{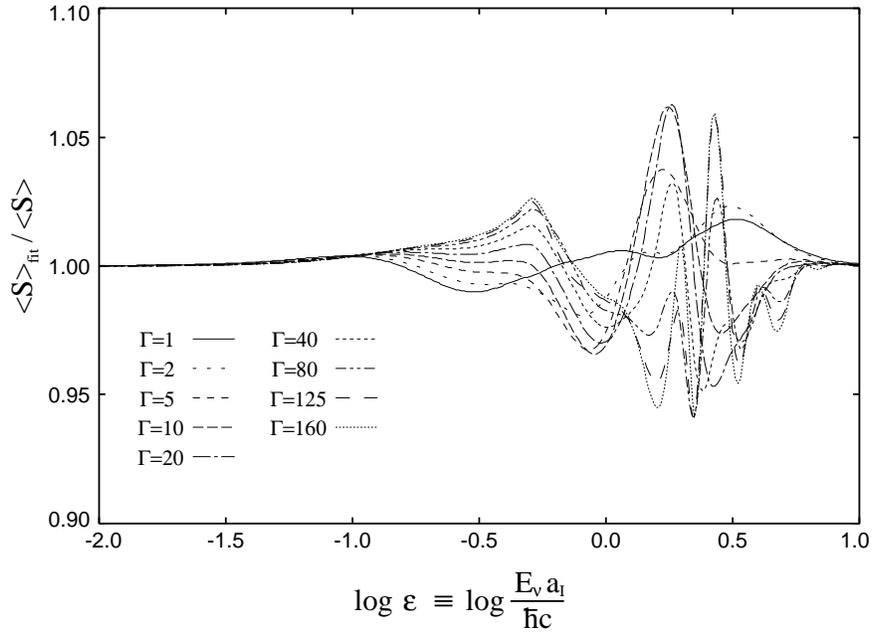}
\caption{The ratio of the present fitting formula value to our numerical result $<S>_{fit}/<S>$.}
\end{figure}

\end{document}